\documentclass[12pt]{article}

\usepackage{amsmath,amsfonts,amssymb,color,epsfig,graphics,graphicx,latexsym,revsymb,theorem,url,verbatim,epstopdf}

\usepackage{hyperref}

\usepackage[cm]{fullpage}

\newtheorem{definition}{Definition}
\newtheorem{proposition}[definition]{Proposition}
\newtheorem{lemma}[definition]{Lemma}

\newtheorem{theorem}[definition]{Theorem}
\newtheorem{corollary}[definition]{Corollary}
\newtheorem{conjecture}[definition]{Conjecture}

\newtheorem{remark}[definition]{Remark}
\newtheorem{example}[definition]{Example}
\newtheorem{question}[definition]{Question}

\def\squareforqed{\hbox{\rlap{$\sqcap$}$\sqcup$}}
\def\qed{\ifmmode\squareforqed\else{\unskip\nobreak\hfil
\penalty50\hskip1em\null\nobreak\hfil\squareforqed
\parfillskip=0pt\finalhyphendemerits=0\endgraf}\fi}
\def\endenv{\ifmmode\;\else{\unskip\nobreak\hfil
\penalty50\hskip1em\null\nobreak\hfil\;
\parfillskip=0pt\finalhyphendemerits=0\endgraf}\fi}
\newenvironment{proof}{\noindent \textbf{{Proof.~} }}{\qed}
\def\Dbar{\leavevmode\lower.6ex\hbox to 0pt
{\hskip-.23ex\accent"16\hss}D}
\makeatletter
\def\url@leostyle{%
  \@ifundefined{selectfont}{\def\UrlFont{\sf}}{\def\UrlFont{\small\ttfamily}}}
\makeatother
\def\bcj{\begin{conjecture}}
\def\ecj{\end{conjecture}}
\def\bcr{\begin{corollary}}
\def\ecr{\end{corollary}}
\def\bd{\begin{definition}}
\def\ed{\end{definition}}
\def\bea{\begin{eqnarray}}
\def\eea{\end{eqnarray}}
\def\bem{\begin{enumerate}}
\def\eem{\end{enumerate}}
\def\bex{\begin{example}}
\def\eex{\end{example}}
\def\bim{\begin{itemize}}
\def\eim{\end{itemize}}
\def\bl{\begin{lemma}}
\def\el{\end{lemma}}
\def\bpf{\begin{proof}}
\def\epf{\end{proof}}
\def\bpp{\begin{proposition}}
\def\epp{\end{proposition}}
\def\bqu{\begin{question}}
\def\equ{\end{question}}
\def\br{\begin{remark}}
\def\er{\end{remark}}
\def\bt{\begin{theorem}}
\def\et{\end{theorem}}

\def\btb{\begin{tabular}}
\def\etb{\end{tabular}}

\newcommand{\nc}{\newcommand}

\def\a{\alpha}
\def\b{\beta}
\def\g{\gamma}

\def\z{\zeta}

\def\r{\rho}

\def\ps{\psi}

\def\G{\Gamma}
\def\D{\Delta}
\def\T{\Theta}

 \nc{\bbA}{\mathbb{A}} \nc{\bbB}{\mathbb{B}} \nc{\bbC}{\mathbb{C}}
 \nc{\bbD}{\mathbb{D}} \nc{\bbE}{\mathbb{E}} \nc{\bbF}{\mathbb{F}}
 \nc{\bbG}{\mathbb{G}} \nc{\bbH}{\mathbb{H}} \nc{\bbI}{\mathbb{I}}
 \nc{\bbJ}{\mathbb{J}} \nc{\bbK}{\mathbb{K}} \nc{\bbL}{\mathbb{L}}
 \nc{\bbM}{\mathbb{M}} \nc{\bbN}{\mathbb{N}} \nc{\bbO}{\mathbb{O}}
 \nc{\bbP}{\mathbb{P}} \nc{\bbQ}{\mathbb{Q}} \nc{\bbR}{\mathbb{R}}
 \nc{\bbS}{\mathbb{S}} \nc{\bbT}{\mathbb{T}} \nc{\bbU}{\mathbb{U}}
 \nc{\bbV}{\mathbb{V}} \nc{\bbW}{\mathbb{W}} \nc{\bbX}{\mathbb{X}}
 \nc{\bbZ}{\mathbb{Z}}

 \nc{\bA}{{\bf A}} \nc{\bB}{{\bf B}} \nc{\bC}{{\bf C}}
 \nc{\bD}{{\bf D}} \nc{\bE}{{\bf E}} \nc{\bF}{{\bf F}}
 \nc{\bG}{{\bf G}} \nc{\bH}{{\bf H}} \nc{\bI}{{\bf I}}
 \nc{\bJ}{{\bf J}} \nc{\bK}{{\bf K}} \nc{\bL}{{\bf L}}
 \nc{\bM}{{\bf M}} \nc{\bN}{{\bf N}} \nc{\bO}{{\bf O}}
 \nc{\bP}{{\bf P}} \nc{\bQ}{{\bf Q}} \nc{\bR}{{\bf R}}
 \nc{\bS}{{\bf S}} \nc{\bT}{{\bf T}} \nc{\bU}{{\bf U}}
 \nc{\bV}{{\bf V}} \nc{\bW}{{\bf W}} \nc{\bX}{{\bf X}}
 \nc{\bZ}{{\bf Z}}

\nc{\cA}{{\cal A}} \nc{\cB}{{\cal B}} \nc{\cC}{{\cal C}}
\nc{\cD}{{\cal D}} \nc{\cE}{{\cal E}} \nc{\cF}{{\cal F}}
\nc{\cG}{{\cal G}} \nc{\cH}{{\cal H}} \nc{\cI}{{\cal I}}
\nc{\cJ}{{\cal J}} \nc{\cK}{{\cal K}} \nc{\cL}{{\cal L}}
\nc{\cM}{{\cal M}} \nc{\cN}{{\cal N}} \nc{\cO}{{\cal O}}
\nc{\cP}{{\cal P}} \nc{\cQ}{{\cal Q}} \nc{\cR}{{\cal R}}
\nc{\cS}{{\cal S}} \nc{\cT}{{\cal T}} \nc{\cU}{{\cal U}}
\nc{\cV}{{\cal V}} \nc{\cW}{{\cal W}} \nc{\cX}{{\cal X}}
\nc{\cZ}{{\cal Z}}

\nc{\hA}{{\hat{A}}} \nc{\hB}{{\hat{B}}} \nc{\hC}{{\hat{C}}}
\nc{\hD}{{\hat{D}}} \nc{\hE}{{\hat{E}}} \nc{\hF}{{\hat{F}}}
\nc{\hG}{{\hat{G}}} \nc{\hH}{{\hat{H}}} \nc{\hI}{{\hat{I}}}
\nc{\hJ}{{\hat{J}}} \nc{\hK}{{\hat{K}}} \nc{\hL}{{\hat{L}}}
\nc{\hM}{{\hat{M}}} \nc{\hN}{{\hat{N}}} \nc{\hO}{{\hat{O}}}
\nc{\hP}{{\hat{P}}} \nc{\hR}{{\hat{R}}} \nc{\hS}{{\hat{S}}}
\nc{\hT}{{\hat{T}}} \nc{\hU}{{\hat{U}}} \nc{\hV}{{\hat{V}}}
\nc{\hW}{{\hat{W}}} \nc{\hX}{{\hat{X}}} \nc{\hZ}{{\hat{Z}}}

\nc{\hn}{{\hat{n}}}

\def\dim{\mathop{\rm Dim}}
\def\End{\mathop{\rm End}}
\def\Hom{\mathop{\rm Hom}}

\def\max{\mathop{\rm max}}

\def\tr{\mathop{\rm Tr}}

\def\ox{\otimes}

\def\pars{\partial\cS}

\def\su{\subset}
\def\sue{\subseteq}

\newcommand{\bra}[1]{\langle#1|}
\newcommand{\ket}[1]{|#1\rangle}
\newcommand{\proj}[1]{| #1\rangle\!\langle #1 |}
\newcommand{\ketbra}[2]{|#1\rangle\!\langle#2|}
\newcommand{\braket}[2]{\langle#1|#2\rangle}

\newcommand{\abs}[1]{|#1|}

\newcommand{\jmp}{J. Math. Phys.}

\newcommand{\pra}{Phys. Rev. A~}
\newcommand{\prl}{Phys. Rev. Lett.}

\def\Dbar{\leavevmode\lower.6ex\hbox to 0pt
{\hskip-.23ex\accent"16\hss}D}

\begin{document}
\title{Dimension formula for induced maximal faces of separable states and genuine entanglement}

\author{ Lin Chen $^{1,2}$, Dragomir {\v{Z} \Dbar}okovi{\'c}$^{3,4}$\\
\small ${}^{1}$ School of Mathematics and Systems Science, Beihang University, Beijing 100191, China \\
\small ${}^{2}$ International Research Institute for Multidisciplinary Science, Beihang University, Beijing 100191, China \\
\small ${}^{3}$ Institute for Quantum Computing, University of
Waterloo, Waterloo, Ontario, Canada\\
\small ${}^{4}$ Department of Pure Mathematics, University of
Waterloo, Waterloo, Ontario, Canada\\
}

\date{\today}


\maketitle

\begin{abstract}
The normalized separable states of a finite-dimensional multipartite quantum system, represented by its Hilbert space $\cH$, form a closed convex set $\cS_1$. The set $\cS_1$ has two kinds of faces, induced and non-induced. An induced face, $F$, has the form $F=\Gamma(F_V)$, where $V$ is a subspace of $\cH$, $F_V$ is the set of $\rho\in\cS_1$ whose range is contained in 
$V$, and $\Gamma$ is a partial transposition operator. Such $F$ is a maximal face if and only if $V$ is a hyperplane. We give a simple formula for the dimension of any induced maximal face. We also prove that the maximum dimension of induced maximal faces is equal to $d(d-2)$ where $d$ is the dimension of $\cH$. The 
equality $\dim\Gamma(F_V)=d(d-2)$ holds if and only if $V^\perp$ is spanned by a genuinely entangled vector.
\end{abstract}

\tableofcontents

\section{Introduction}\label{sec1}

Let $\cH=\cH_1\otimes\cH_2\otimes\cdots\otimes\cH_n$ be the
complex Hilbert space of a finite-dimensional $n$-partite quantum
system. We denote by $d_i$ the dimension of $\cH_i$, and so
$d:=\prod d_i$ is the dimension of $\cH$. We assume that each 
$d_i>1$. 
A vector $\ket{x}\in\cH$ is {\em normalized} if $\|x\|=1$. We denote by $H$ the space of Hermitian operators $\r$ on $\cH$. Note that $H$ is a real vector space of dimension $d^2$. The mixed quantum states of this  system are represented by their density matrices, i.e., operators $\r\in H$ which are positive semidefinite $(\r\ge0)$ and have unit trace $(\tr \r=1)$. For convenience, we often work with non-normalized states, i.e., Hermitian operators $\r$ such that $\r\ge0$ and $\r\ne0$. It will be clear from the context whether we require the states to be normalized. We denote by $\cR(\r)$ the range of a linear operator $\r$.

We assume that an orthonormal basis is fixed in each $\cH_i$
and we use the standard notation $\ket{0},\ldots,\ket{d_i-1}$
for the corresponding basis vectors. We write $\End V$ for the algebra of linear operators on a complex vector space $V$. 
The operation of transposition applied only to the $i$th tensor factor of $\End\cH=\ox_{i=1}^n \End\cH_i$ will be denoted by $\G_i$. (The transposition is with respect to the basis fixed above.) We denote by $\Theta$ the abelian group of order $2^n$ generated by the $\G_i$s. We refer to the elements of $\T$ as the {\em partial transposition operators}. Thus if $\r$ is a state on $\cH$, then $\G_i(\r)$ is the $i$th partial transpose of $\r$. We recall a definition from 
\cite{cd14}: an operator $\r\in H$ is {\em full} if $\G(\r)$ has rank $d$ for all $\G\in\T$. 

A {\em product vector} is a nonzero vector of the form $\ket{x}=\ket{x_1}\ox\cdots\ox\ket{x_n}$ where $\ket{x_i}\in\cH_i$. We shall write this product vector also as $\ket{x_1,\ldots,x_n}$. 
A {\em pure product state} is a state $\r$ of the form $\r=\proj{x}$ where $\ket{x}$ is a product vector. The product vectors 
$\ket{i_1,i_2,\ldots,i_n}$, $0\le i_k<d_k$, form an orthonormal (o.n.) basis of $\cH$. A quantum state $\r$ is {\em separable} if it is a sum of pure product states, i.e., $\r=\sum_{k=1}^l \proj{z_k}$, 
where the $\ket{z_k}$ are product vectors. A non-separable state is also called an \emph{entangled state}. Quantum entanglement has been used to realize many tasks, such as quantum teleportation \cite{bbc93}, cryptography \cite{bb84} and dense coding, that surpass their classical counterparts both theoretically and experimentally. Deciding whether a given state is entangled turns out to be hard. This problem has been solved only for the Hilbert spaces of dimension $2\times2$ and $2\times3$ \cite{hhh96}. Thus it is important to understand the properties of the set $\cS_1$ of normalized separable states. In this paper, we will investigate the faces of $\cS_1$. The full separable states that lie on the boundary, $\pars_1$, of $\cS_1$ are of special interest.

We denote by $\cD_1$ and $\cD$ the set of normalized and
non-normalized states, respectively. Thus 
$\cD_1=\{\r\in\cD:\tr\r=1\}$ is a compact convex subset of 
the affine hyperplane of $H$ defined by the equation $\tr\r=1$.
The faces of $\cD_1$ are parametrized by vector subspaces
$V\sue\cH$ \cite[section II]{as10}. The face $\tilde{F}_V$ that corresponds to $V$ consists of all states $\r\in\cD_1$ such that $\cR(\r)\sue V$. The intersection 
 \bea \label{eq:FaceF_V}
F_V:=\tilde{F}_V\cap\cS_1=\{\r\in\cS_1:\cR(\r)\sue V\}
 \eea
is a face (possibly empty) of $\cS_1$. We say that the face $F_V$ is {\em associated} to $V$. The following basic fact was proved recently \cite[Theorem 15]{cd14}.

\bt \label{thm:MaxFace}
For a subspace $V\sue\cH$, the associated face $F_V$ of $\cS_1$ is maximal if and only if $V$ is a hyperplane of $\cH$. 
\et

As in \cite{cd14} we enlarge the collection of faces of type 
$F_V$ by using the group $\T$. 

\bd 
A face $F$ of $\cS_1$ is {\em induced} if $F=\G(F_V)$ for some
vector subspace $V\sue\cH$ and some partial transposition 
$\G\in\T$.
\ed

We warn the reader that our general definition above is different from the one adopted (for the bipartite case) by H.-S. Choi and S.-H, Kye \cite{CK2012}.

The boundary of $\cS_1$ is the union of all maximal faces and 
in order to describe geometrically the set $\cS_1$ one needs 
the description of the maximal faces. The induced maximal faces 
have the form $\G(F_V)$ where $V$ is a hyperplane of $\cH$ and 
$\G\in\T$. The non-induced faces are more elusive. It was shown in \cite{cd14} that there exist maximal faces of $\cS_1$ which are not induced. Explicit examples of non-induced maximal faces of $\cS_1$ are presently known in several systems, notably in 
$3\ox3$ and $2\ox4$ systems. A remarkable family of non-induced faces of $\cS_1$ in the case of two qutrits was constructed recently in \cite{HaKye}. Each of these faces is a 9-dimensional simplex. Moreover, each of them is the intersection of two maximal faces \cite{Kye}. For more details see section \ref{sec:max}.

It is certainly of interest to know the maximum dimension of maximal faces of $\cS_1$. In this regard, we propose the following conjecture.

\bcj \label{cj:max}
For any finite dimensional quantum system we have
\bea \label{eq:Max}
\max_{F,~{\rm maximal~face}} \dim F =
\max_{V\subset\cH,~{\rm hyperplane}} \dim F_V.
\eea
\ecj

Since $\dim \G(F_V)=\dim F_V$ for any $\G\in\T$ and any vector subspace $V\sue\cH$, the rhs is equal to the maximum dimension taken over all induced maximal faces of $\cS_1$. 
The lhs is obviously greater than or equal to the rhs. 
We shall prove that the rhs is equal to $d(d-2)$ (see Corollary \ref{cr:dim-Va}). 
The number $d(d-2)$ arises very naturally. Indeed, 
$\dim \pars_1 = d^2-2$ and induced maximal faces are parameterized by the complex projective space
whose real dimension is  $2(d-1)$. Therefore, the number
$(d^2-2) - 2(d-1) = d(d-2)$ is expected. The genuinely 
entangled vectors are generic and the maximal faces induced 
by the hyperplanes orthogonal to such vectors indeed have 
the expected dimension.

The lhs of \eqref{eq:Max} is known only when $d\le6$, and in these cases the conjectured equality holds \cite{cd14}. 
See section \ref{sec:max} for additional evidence in support of the conjecture.

If $\ket{\a}\in\cH$ is a product vector and $V=\ket{\a}^\perp$ then we have shown in \cite[Proposition 17]{cd14} that 
\bea \label{eq:Dim-1}
\dim F_V=d^2-1-\prod_{i=1}^n (2d_i-1).
\eea

The main result of the paper, Theorem \ref{thm:dim-V},  generalizes this formula. It provides a method for computing the dimension of any induced maximal face in any multipartite quantum system. It turns out that $\dim F_V$, $V$ a hyperplane, depends only on the full tensor factorization of the vector $\ket{\a}$ orthogonal to $V$. Corollary \ref{cr:dim-Va} shows that the genuine entanglement \cite{ckw00} from quantum information is linked to the geometry of faces of $\cS_1$.  We say that $\ket{\a}$ is {\it genuinely entangled (g.e.)} if it is entangled for all bipartitions of $\cH$. This corollary states that $\ket{\a}$ is g.e. if and only if $\dim F_V =d(d-2)$. For instance, when $\ket{\a}$ is the three-qubit Greenberger-Horne-Zeilinger (GHZ) state (which is g.e. \cite{ghz}) then $\dim F_V=48$.
Note that, according to the above definition, in the case $n=1$ each nonzero vector is g.e. by default since the system admits no bipartitions. (This makes sense only mathematically, as entanglement physically exists only when $n>1$.) 

From the viewpoint of experiments, the genuine entanglement is easier to detect than the non-genuine one \cite{hhr05}. Most extensively studied quantum states are g.e. such as the multiqubit GHZ, W and Dicke states, and the 4-qubit L and M states \cite{gw10}. They indeed play essential role in the implementation of many quantum-information tasks. In particular, entanglement witnesses detecting genuine multiqubit entanglement have been realized recently in experiments \cite{twk09}. Many g.e. states can be constructed by using combinatorial designs \cite{gz14}.

The rest of the paper is organized as follows. In Sec. \ref{sec:dim} we prove our main result, Theorem \ref{thm:dim-V}, and 
derive two important corollaries \ref{cr:dim-Va} and \ref{cr:dim-Vb}. In Sec. \ref{sec:Theta}, we study the action of $\Theta$ on induced maximal faces. The main technical result is Theorem \ref{thm:face}. The two corollaries \ref{cor:d-1} and \ref{cor:Theta'} describe the action of $\T$ in more details. 
In particular, it is shown that if $V:=\ket{\a}^\perp$ then the face $F_V$ and the state $\proj{\a}$ have the same stabilizer in $\T$. In section \ref{sec:max} we examine two known infinite families of non-induced maximal faces (in $3\ox3$ and $2\ox4$) 
and show that their dimensions are less than $d(d-2)$. 
We conclude in Sec. \ref{sec:conclusion}.

\section{Dimension formula}
\label{sec:dim}

We shall compute the dimension of the maximal face $F_V$ where 
$V\subseteq\cH$ is any hyperplane.
We shall need the following elementary lemma and its corollary. 

\bl \label{le:fg=0}
Let $\sum_{i=1}^n f_ig_i^*=0$ where the $f_i$ and the $g_i$ 
are polynomials in independent complex variables 
$z_1,\ldots,z_m$. If the $f_i$ are linearly independent, 
then all $g_i=0$.
\el
\bpf
We use induction on $n$. The case $n=1$ is trivial. Assume that 
$n>1$. Let $\mu$ be a monomial in $z_1,\ldots,z_m$ which occurs in $f_1$. Choose $c\in\bC$ such that $\mu$ does not occur in 
$f_2-cf_1$. We can replace $f_2$ with $f_2-cf_1$ and $g_1$ with 
$g_1+c^*g_2$ while preserving the equality $\sum_{i=1}^n f_ig_i^*=0$. Therefore we may assume that $\mu$ does not occur
in $f_2$. Similarly, we may assume that it does not occur in 
any other $f_i$. By \cite[Lemma 7]{cd14} we must have $g_1=0$. By the induction hypothesis, all other $g_i=0$.
\epf

\bcr \label{cor:fg-system}
Let $f_1,\ldots,f_p$ and $g_1,\ldots,g_q$ be polynomials in 
independent complex variables $z_1,\ldots,z_m$. If both 
$f_1,\ldots,f_p$ and $g_1,\ldots,g_q$ are linearly independent, 
then also the products $f_ig^*_j$ are linearly independent.
\ecr
\bpf 
Assume that $\sum_{i,j} c_{ij}f_ig^*_j=0$ for some 
$c_{ij}\in\bC$. Thus 
$\sum_i f_i\left(\sum_j c_{ij}^*g_j\right)^*=0$ and the lemma 
implies that $\sum_j c_{ij}^*g_j=0$ for each $i$. As the $g_j$ 
are linearly independent, all $c_{ij}=0$.
\epf

We recall another important fact which we use in our proofs.
Let $H$ be the real subspace of the algebra $\End \cH$  consisting of all Hermitian operators. 
As usual we identify $\End \cH$ with the space $M$ of 
$d\times d$ complex matrices. Let $T$ be the real subspace of $M$ consisting of all upper triangular matrices with real diagonal elements. The map $H\to T$ which sends a Hermitian matrix to its upper triangular part is an isomorphism of real vector spaces. For any subspace $L$ of $H$, we have $\dim L = \dim L'$ where $L'$ denotes the image of $L$ by this isomorphism. 

It is immediate from our definition of genuinely entangled states that each nonzero vector $\ket{\a}\in\cH$, after a permutation of the parties, can be written as the tensor product 
\bea \label{eq:ge-decomp}
\ket{\a}=\ket{\a'_1}\ox\ket{\a'_2}\ox\cdots\ox\ket{\a'_m}
=\ket{\a'_1,\a'_2,\ldots,\a'_m},
\eea
where each $\ket{\a'_i}$ is g.e.
The tensor factors $\ket{\a'_i}$ are unique up to scalar factors and a permutation, and we shall refer to them as the
{\it g.e. components} of $\ket{\a}$. We also say that 
\eqref{eq:ge-decomp} is a {\it g.e. decomposition} of $\ket{\a}$. Using this terminology, we state our main theorem which generalizes our previous result \cite[Proposition 17]{cd14}.

\bt
\label{thm:dim-V}
Let $V\su\cH$ be any hyperplane and let $\ket{\a}\in V^\perp$ be 
a nonzero vector. Let \eqref{eq:ge-decomp} be the g.e. 
decomposition of $\ket{\a}$ with
$\ket{\a'_i}\in\cH'_i:=\cH_{n_1+\cdots+n_{i-1}+1}\ox\cdots\ox
\cH_{n_1+\cdots+n_i}$ and $n=n_1+n_2+\cdots+n_m$, $n_i\ge1$.
Then 
\bea \label{eq:dim-V=}
\dim F_V = d^2-1-\prod_{i=1}^m (2d_i'-1),
\eea
where $d'_i=\dim\cH'_i$.
\et
\bpf
We can view $\cH$ also as a quantum system consisting of $m$ parties $\cH'_i$, $i=1,\ldots,m$. In that case we shall denote it by $\cH'$ and its set of (normalized) separable states by 
$\cS'_1$. Note that $\cS_1\subseteq\cS'_1$, and if $m<n$ then the  inclusion is proper. 
Thus, if $F'_V$ is the face of $\cS'_1$ associated to $V$, then \eqref{eq:FaceF_V} implies that $F_V\subseteq F'_V$ and so 
$\dim F_V\le\dim F'_V$. Since $\ket{\a}$ is a product vector in $\cH'$, by applying the formula \eqref{eq:Dim-1}, we obtain that 
\bea \label{eq:prva-nej}
\dim F'_V = d^2-1-\prod_{i=1}^m (2d_i'-1).
\eea

It remains to prove that $\dim F_V = \dim F'_V$. We use the 
induction on $m$. The crucial case is $m=1$, i.e., when 
$\ket{\a}$ is genuinely entangled. In that case we know that 
$\dim F_V\le\dim F'_V=d(d-2)$.

Assume that $\dim F_V<d(d-2)$ and let 
\bea
\label{eq:vekta}
\ket{\a}=\sum^{d_1-1}_{j_1=0} \sum^{d_2-1}_{j_2=0} \cdots
\sum^{d_{n}-1}_{j_{n}=0}
a_{j_1,j_2,\ldots,j_n}\ket{j_1,j_2,\ldots,j_n}.
\eea
Without any loss of generality we may assume that the coefficient $a_{d_1-1,d_2-1,\ldots,d_n-1}\ne0$. Let $\ket{\z}:=\ket{\z_1,\z_2,\ldots,\z_n}\in\cH$ be any product vector with 
\bea \label{eq:prodz}
\ket{\z_i}=\sum^{d_i-1}_{j_i=0} z_{i,j_i}\ket{j_i},
\quad i=1,\ldots,n.
\eea
The inner product $P:=\braket{\a}{\z}$ is a polynomial of degree 
$n$ which is multilinear in the $n$ sets of complex variables
$\{z_{i,j_i}:~ j_i=0,1,\ldots,d_i-1\}$, $i=1,\ldots,n$. 
In order to write this polynomial explicitly (and for later use) it is convenient to introduce the abbreviations
\bea \label{eq:def-f}
f_{j_1,j_2,\ldots,j_k}=z_{1,j_1}z_{2,j_2}\cdots z_{k,j_k},
\quad 1\le k\le n.
\eea
Then we have
\bea
\label{eq:az}
P=\sum^{d_1-1}_{j_1=0} \sum^{d_2-1}_{j_2=0} \cdots
\sum^{d_{n}-1}_{j_{n}=0}
a^*_{j_1,j_2,\ldots,j_n} f_{j_1,j_2,\ldots,j_n}.
\eea
Let $L\sue H$ be the real span of all pure product states 
\bea
\label{eq:square}
\proj{\z}=\sum_{(j_1,j_2,\ldots,j_n)}\sum_{(k_1,k_2,\ldots,k_n)}
f_{j_1,j_2,\ldots,j_n} f^*_{k_1,k_2,\ldots,k_n}
\ketbra{j_1,j_2,\ldots,j_n}{k_1,k_2,\ldots,k_n},
\eea
where the complex varables $z_{i,j_i}$ are subject only to the 
constraint $P=0$ (which guarantees that $\ket{\z}\in V$). Let us denote by $\cX$ the hypersurface in 
$\cH_1\times\cH_2\times\cdots\times\cH_n$ defined by the 
equation $P=0$. Here we consider $H$ as 
the space of $d\times d$ Hermitian matrices whose rows and columns are indexed by the $n$-tuples $(j_1,j_2,\ldots,j_n)$ with 
$j_k\in\{0,1,\ldots,d_k-1\}$. These $n$-tuples are linearly ordered so that $(j_1,j_2,\ldots,j_n)<(k_1,k_2,\ldots,k_n)$ 
holds if and only if $j_i=k_i$ for $i<s$ and $j_s<k_s$ for 
some index $s$. 

Since $a_{d_1-1,d_2-1,\ldots,d_n-1}\ne0$, the monomial 
$f_{d_1-1,d_2-1,\ldots,d_n-1}$ (when restricted to $\cX$) is a linear combination with constant coefficients of $d^2-1$ other monomials. Thus, we can eliminate this monomial (and its complex conjugate) from \eqref{eq:square}. Then this equation can be rewritten as
\bea \label{eq:zeta}
\proj{\z} = \sum f_{j_1,\ldots,j_n} f^*_{k_1,\ldots,k_n} 
\r_{j_1,\ldots,j_n;k_1,\ldots,k_n},
\eea
where $\r^\dag_{j_1,\ldots,j_n;k_1,\ldots,k_n}=
\r_{k_1,\ldots,k_n;j_1,\ldots,j_n}$ and neither 
$(j_1,\ldots,j_n)$ nor $(k_1,\ldots,k_n)$ is equal to 
$(d_1-1,\ldots,d_n-1)$. Thus this sum has $(d-1)^2$ terms. 
Moreover, the $(d-1)^2$ matrices 
$\r_{j_1,\ldots,j_n;k_1,\ldots,k_n}$ are linearly independent.
The coefficient of $\r_{j_1,\ldots,j_n;k_1,\ldots,k_n}$ is real (and nonnegative) if $(j_1,\ldots,j_n)=(k_1,\ldots,k_n)$ 
and the other coefficients occur in complex conjugate pairs. 

As $\dim F_V<d(d-2)$, we have $\dim L<(d-1)^2$. We have  
$P=P_0 z_{n,d_n-1}+P_1$, where 
\bea
\label{eq:w}
&&
P_0 := 
\sum^{d_1-1}_{j_1=0} 
\cdots 
\sum^{d_{n-1}-1}_{j_{n-1}=0} 
a^*_{j_1,\ldots,j_{n-1},d_n-1} f_{j_1,\ldots,j_{n-1}},
\\
\label{eq:x}
&&
P_1 :=
\sum^{d_1-1}_{j_1=0} 
\cdots 
\sum^{d_{n-1}-1}_{j_{n-1}=0} 
\sum^{d_{n}-2}_{j_{n}=0}
a^*_{j_1,\ldots,j_{n-1},j_n} f_{j_1,\ldots,j_{n-1},j_n}.
\eea
As in the previous paragraph, we know that $L$ is spanned by the matrices \eqref{eq:zeta} with the restriction on the indices 
$j_i$ and $k_i$ specified there. Let $L'$ be the subspace of $L$ spanned by all matrices \eqref{eq:zeta} with the additional constraint $P_0\ne0$ on the variables $z_{i,j_i}$. Obviously, we have $\dim L' \le \dim L<(d-1)^2$.

Assume that the polynomial $P$ is irreducible. 
By multiplying  \eqref{eq:zeta} with $|P_0|^2$, we obtain the equation 
\bea \label{eq:W.zeta}
|P_0|^2\cdot\proj{\z} = \sum (P_0 f_{j_1,\ldots,j_n})
(P_0 f_{k_1,\ldots,k_n})^* 
\r_{j_1,\ldots,j_n;k_1,\ldots,k_n}.
\eea
The variable $z_{n,d_n-1}$ can be eliminated from the rhs by 
using the equation $P_0z_{n,d_n-1}=-P_1$. For instance, if $j_n=d_n-1$ 
we can replace $P_0z_{n,j_n}$ with $-P_1$. After this elimination, the coefficients of $\r_{j_1,\ldots,j_n;k_1,\ldots,k_n}$ on the rhs will have the form $gh^*$, where 
\bea
g,h\in\Phi &:=& \{ P_0 f_{k_1,\ldots,k_{n-1},k_n}: k_n\ne d_n-1\} 
\cup \notag  \\
\label{eq:fk1}
&&
\{-P_1 f_{k_1,\ldots,k_{n-1}}: \exists i<n, ~k_i<d_i-1 \}.
\eea

We claim that the $d-1$ polynomials in $\Phi$ are linearly independent. Otherwise we have an identity $P_0Q_1=P_1Q_0$ 
where $Q_1$ is a nonzero multilinear polynomial of degree $n$
and the monomial $f_{d_1-1,\ldots d_{n-1}-1}$ does not occur in $Q_0$. Since $P$ is irreducible, $P_0$ and $P_1$ are 
relatively prime. Hence, we may assume that $P_0=Q_0$. But this is impossible since $f_{d_1-1,\ldots d_{n-1}-1}$ occurs in $P_0$. Thus, our claim is proved.

By Corollary \ref{cor:fg-system}, the $(d-1)^2$ products $gh^*$ with $g,h\in\Phi$ are also linearly independent. As the matrices 
$\r_{j_1,\ldots,j_n;k_1,\ldots,k_n}$ are linearly independent, 
it follows from \eqref{eq:W.zeta} that $\dim L' = (d-1)^2$. This contradiction implies that $P$ is reducible, i.e., it has a nontrivial factorization $P=P'P''$, where $P'$ and $P''$ are homogeneous polynomials of degree $m$ and $n-m$, respectively. Since $P$ is multilinear, the same is true for $P'$ and $P''$. Hence, after permuting the $\cH_i$, we may assume that $P'$ depends only on the variables $z_{i,j_i}$ with $i\le m$, and $P''$ depends only on the 
$z_{i,j_i}$ with $i>m$. This means that 
$\ket{\a}=\ket{\a'}\ox\ket{\a''}$ with 
$\ket{\a'}\in\cH_1\ox\cdots\ox\cH_m$ and 
$\ket{\a''}\in\cH_{m+1}\ox\cdots\ox\cH_n$, i.e., $\ket{\a}$ 
is not genuinely entangled. Thus, we have finally reached a contradiction. Hence, we conclude that 
$\dim F_V=\dim F'_V=d(d-2)$ when $m=1$.

Now let $m>1$. Let us introduce the quantum system 
$\hat{\cH}=\cH'_1\ox\cdots\ox\cH'_{m-1}$, which we consider 
as a subsystem of $\cH'$. We set 
$\ket{\hat{\a}}=\ket{\a'_1,\ldots,\a'_{m-1}}$, 
$\hat{V}=\ket{\hat{\a}}^\perp\subset\hat{\cH}$ and 
$V'_i=\ket{\a'_i}^\perp\subset\cH'_i$. 
Denote the space of Hermitian operators on $\hat{\cH}$ and 
$\cH'_i$ by $\hat{H}$ and $H'_i$, respectively. By the induction hypothesis we have 
$\dim F_{\hat{V}}=\dim F'_{\hat{V}}$ and, by \eqref{eq:Dim-1}, 
we have 
$\dim F'_{\hat{V}}=(d/d'_m)^2-1-\prod_{i=1}^{m-1} (2d_i'-1)$.

Let $\ket{\z}$ be any product vector in $V$, and note that 
$\ket{\z}\in\hat{\cH}\ox V'_m$ or $\ket{\z}\in\hat{V}\ox\cH'_m$. 
Let $L\sue H$ be the real span of all $\proj{\z}$. 
Similarly, let $\hat{L}\subset\hat{H}$ be the real span of 
$\proj{\xi}$ over all product vectors $\ket{\xi}\in\hat{L}$ orthogonal to $\ket{\hat{\a}}$. Finally, let $L'_i$ be the real span of $\proj{\eta}$ over all product vectors $\ket{\eta}\in\cH'_i$ orthogonal to $\ket{\a'_i}$. 
Since we have already handled the case $m=1$, we know that 
$\dim L'_m=(d'_m-1)^2$. 

The $\proj{\z}$ with $\ket{\z}\in\hat{\cH}\ox V'_m$ span the space $\hat{H}\ox_\bR L'_m$, and those with 
$\ket{\z}\in\hat{V}\ox\cH'_m$ span the space 
$\hat{L}\ox_\bR H'_m$. Consequently, we have
$L=\hat{L}\ox_\bR H'_m + \hat{H}\ox_\bR L'_m$.
Since $\dim L'_m=(d'_m-1)^2$ 
and $\dim \hat{L}=\dim F'_{\hat{V}}+1=
(d/d'_m)^2-\prod_{i=1}^{m-1} (2d_i'-1)$,
it follows that 

\bea
\dim L &=& \dim \hat{L}\ox_\bR H'_m + \dim \hat{H}\ox_\bR L'_m
-\dim \hat{L}\ox_\bR L'_m \notag \\
&=& (d'_m)^2 \dim \hat{L} + (d/d'_m)^2 (d'_m-1)^2
-(d'_m-1)^2 \dim \hat{L} 
\notag\\
&=& d^2-\prod_{i=1}^m (2d_i'-1).
\eea
This completes the proof of the induction step, and of the theorem.
\epf

We remark that $\prod_{i=1}^m (2d'_i-1)\ge2d-1$ and that equality holds only if $m=1$. Thus we have the following corollary.

\bcr
\label{cr:dim-Va}
For any hyperplane $V=\ket{\a}^\perp\subset\cH$ we have 
$\dim F_V\le d(d-2)$ and equality holds if and only if $\ket{\a}$ 
is genuinely entangled.
\ecr

Similarly one can prove the following corollary.
\bcr
\label{cr:dim-Vb}
If we drop in Theorem \ref{thm:dim-V} the hypothesis that each 
$\ket{\a'_i}$ is g.e., then 
\bea
\label{eq:dim-V}
\dim F_V \le d^2-1-\prod_{i=1}^m (2d_i'-1),
\eea
and the equality holds if and only if each $\ket{\a'_i}$ is g.e.
\ecr

\bex {
\label{ex:bipartite}
\rm
Let us consider the  bipartite case $d_1\ox d_2$ $(n=2)$ and let $V=\ket{\a}^\perp$ where $\ket{\a}\in\cH$ is a vector of Schmidt rank $s>0$. If $s>1$ then the pure state $\ket{\a}$ is genuinely entangled and so $m=1$ and $d'_1=d$. Thus \eqref{eq:dim-V} gives that $\dim F_V = d(d-2)$. On the other hand, if $s=1$ then $m=2$,  
$d'_1=d_1$ and $d'_2=d_2$ and so 
$\dim F_V=d^2-1-(2d_1-1)(2d_2-1)$. }
\eex

In general, if $m=n$ in Theorem \ref{thm:dim-V} then the formula 
\eqref{eq:dim-V=} reduces to \eqref{eq:Dim-1}.

\section{Action of $\Theta$ on induced maximal faces} \label{sec:Theta}

For any subset $S\subseteq\{1,2,\ldots,n\}$ we set 
$\G_S=\prod_{i\in S} \G_i$. Moreover, for any $\G\in\T$ there is a unique $S$ such that $\G=\G_S$. Note that if $S=\emptyset$ then 
$\G_S$ is the identity map, and if $S=\{1,2,\ldots,n\}$ then $\G_S$ is the transposition map on $\End \cH$. 
For any vector subspace $V\subseteq\cH$ we denote by $P_V$ the set of all product vectors in $V$. 

Let us introduce the general notion of partial conjugates.
\bd \label{def:S-conj}
If $\ket{\z}=\ket{\z_1,\z_2,\ldots,\z_n}\in\cH$ is a product vector and $S\subseteq\{1,2,\ldots,n\}$, then its
{\em $S$-partial conjugate}, $\ket{\z^{*S}}$, is the product vector 
\bea \label{eq:parc-konj}
\ket{\z^{*S}}=\ket{z_1,z_2,\ldots,z_n},
\eea
where $\ket{z_i}=\ket{\z^*_i}$ if $i\in S$ and 
$\ket{z_i}=\ket{\z_i}$ otherwise.
(Note that this conjugate is well defined only up to a phase factor.) 
\ed

For any subset $X$ of $H$ we set 
$\cR(X)=\sum_{\rho\in X} \cR(\rho)$ and we define the 
{\em rank} of $X$, $r(X)$, to be the dimension of $\cR(X)$. 
This rank can be used to distinguish the induced from the 
non-induced maximal faces of $\cS_1$.
Indeed, if $F$ is a maximal face of $\cS_1$, it was shown in \cite[Proposition 13]{cd14} that $r(F)\ge d-1$ and that $F$ is induced if and only if $r(\G(F))=d-1$ for some $\G\in\T$. 
In particular, for any hyperplane $V\subset\cH$ and any 
$\G\in\T$, the rank of $\G(F_V)$ is either $d$ or $d-1$.
The two cases can be distinguished by the first corollary of the following theorem.

\bt
\label{thm:face}
Let $\ket{\a}\in\cH$ be a nonzero vector, $V=\ket{\a}^\perp$, and let $S=\{1,2,\ldots,k\}$ with $1\le k<n$. 
If $\ket{\a}$ is not separable for the bipartition $S|\bar{S}$, 
then $r(\G_S(F_V))=d$.
\et
\bpf 
Let us define $P^{*S}_V:=\{\ket{\z^{*S}}:\ket{\z}\in P_V\}$.
It is easy to verify that $P^{*S}_V\subseteq\cR(\G_S(F_V))$. 
Hence, in order to prove the theorem it suffices to prove 
that $P^{*S}_V$ spans $\cH$. We shall prove it by 
contradiction.

Assume that $P^{*S}_V$ does not span $\cH$. A product vector 
$\ket{\z}=\ket{\z_1,\z_2,\ldots,\z_n}$ belongs to $V$ if and only if $\braket{\a}{\z}=0$. We shall write $\ket{\a}$ as in 
\eqref{eq:vekta} and the vectors $\ket{\z_i}$ as in 
\eqref{eq:prodz}. We shall also use the abbreviations 
\eqref{eq:def-f}. The inner product $P:=\braket{\a}{\z}$  is given explicitly by  the formula \eqref{eq:az}.
By our assumption, the set $P^{*S}_V$ is contained in some hyperplane of $\cH$. Hence, there exist complex constants 
$b_{j_1,j_2,\ldots,j_n}$ (not all 0) such that 
\bea
\label{eq:b-z}
\sum^{d_1-1}_{j_1=0} \sum^{d_2-1}_{j_2=0} \cdots
\sum^{d_{n}-1}_{j_{n}=0}
b_{j_1,j_2,\ldots,j_n} f^*_{j_1,j_2,\ldots,j_k}
f_{j_{k+1},\ldots,j_n}=0
\eea
holds whenever $P=0$. Let us denote the lhs of the above equation 
by $Q$, it is a polynomial in the variables $z^*_{i,j_i}$, 
$i\le k$, and the variables $z_{i,j_i}$, $i>k$.
This means that $P=0 \Longrightarrow Q=0$. Equivalently, 
\bea
\label{eq:posledica}
|P|^2=0 \Longrightarrow |Q|^2=0.
\eea

Let $P=P_1 P_2\cdots P_r$ and $Q=Q_1 Q_2\cdots Q_s$
be the factorizations into the product of irreducible polynomials. Then we have
\bea
\prod_{i=1}^r |P_i|^2=0 \Longrightarrow 
\prod_{j=1}^s |Q_j|^2=0.
\eea
We view the polynomials $|P_i|^2$ and $|Q_j|^2$ as polynomials in the real and imaginary parts of the complex variables 
$z_{p,q_p}$. As such they have real coefficients and are irreducible over $\bR$. 
The factors $|P_i|^2$ are pairwise non-proportional since they 
depend on different and disjoint sets of variables. 
Consequently, we must have $r=s$ and we may assume that 
$|P_i|^2=|Q_i|^2$ for each $i$. Moreover, we may also assume that  $Q_i=P_i$ or $Q_i=P_i^*$ for $i=1,\ldots,r$.

Note that $P$ is a multilinear polynomial of degree $n$ in the coordinates of the vectors $\ket{\z_i}$, $i=1,\ldots,n$, and
$Q$ is a multilinear polynomial of degree $n$ in the coordinates of the vectors $\ket{\z^*_i}$, $i\le k$ and $\ket{\z_i}$, $i>k$. 
If $n_i$ is the degree of $P_i$, it follows that $P_i$ is a multilinear polynomial in the coordinates of $n_i$ of the vectors 
$\ket{\z_1},\ldots,\ket{\z_n}$. 
Since $\ket{\a}$ is not separable for the bipartition 
$S|\bar{S}$, at least one of the $P_i$, say $P_1$, depends on the coordinates of at least one $\ket{\z_p}$ with $p\le k$ and at least one $\ket{\z_q}$ with $q>k$. 
Consequently, $Q_1$ depends on the vectors $\ket{\z^*_p}$ and 
$\ket{\z_q}$. Hence, $Q_1$ is not a scalar multiple of $P_1$ or $P^*_1$. Thus we have a contradiction.
\epf

The following two corollaries describe the action of $\T$ on the set of induced maximal faces of $\cS_1$.

\bcr \label{cor:d-1}
For any induced maximal face $F$ of $\cS_1$ the following are
equivalent:

(i) $r(F)=d-1$;

(ii) $F=F_V$ for some hyperplane $V\subset\cH$;

(iii) $F=F_{\cR(F)}$.
\ecr
\bpf
As $F$ is a maximal face, we have $r(F)\ge d-1$.

(i) $\Longrightarrow$ (ii) Since $F$ is an induced maximal face, 
we have $F=\G_S(F_V)$ for some hyperplane $V$ and some 
$S\subseteq\{1,2,\ldots,n\}$. Without any loss of generality, we may assume that $S=\{1,2,\ldots,k\}$ for some $k$. Let $\ket{\a}$ be a nonzero vector orthogonal to $V$. By Theorem \ref{thm:face}, we have $\ket{\a}=\ket{\b,\g}$ with 
$\ket{\b}\in\cH_1\ox\cdots\ox\cH_k$ and 
$\ket{\g}\in\cH_{k+1}\ox\cdots\ox\cH_n$. Then $F=F_{V'}$ where 
$V'$ is the hyperplane orthogonal to $\ket{\b^*,\g}$.

(ii) $\Longrightarrow$ (iii) By the definition of $F_V$ we have 
$\cR(F)=\cR(F_V)\subseteq V$. As $r(F)\ge d-1$, it follows that 
$\cR(F)=V$ and so (iii) holds.

(iii) $\Longrightarrow$ (i) Since $F\ne\cS_1$, we have 
$\cR(F)\ne\cH$. Hence, we must have $r(F)=d-1$. 
\epf

\bcr \label{cor:Theta'}
Let \eqref{eq:ge-decomp} be the g.e. decomposition of a nonzero vector $\ket{\a}$, $V=\ket{\a}^\perp$, and let 
$\cH'=\cH'_1\otimes\cH'_2\otimes\cdots\otimes\cH'_m$ 
be the $m$-partite quantum system obtained from $\cH$ by splitting the $n$ parties of $\cH$ 
into $m$ groups as in Theorem \ref{thm:dim-V}. Denote by 
$\T'$ the group of partial transposition operators of the 
quantum system $\cH'$. For $\G\in\T$ we have

(i) $r(\G(F_V))=d-1$ if and only if $\G\in\T'$;

(ii) $\G$ fixes $\proj{\a}$ if and only if $\G\in\T'$ and 
$\ket{\a'_i}$ is real up to a phase factor whenever $\G$ acts on $\End \cH'_i$ non-trivially (i.e., as the transposition map).

(iii) $\G$ fixes $\proj{\a}$ if and only if it fixes $F_V$;
\ecr
\bpf
As (i) and (ii) follow easily from the theorem, we shall prove 
only (iii). By (i) and (ii) we may assume that $\G\in\T'$. After permuting the $\cH'_i$, we may assume that $\G=\G_S$ acts as the transposition operator on $\cH'_1\ox\cdots\ox\cH'_k$ and as identity on $\cH'_{k+1}\ox\cdots\ox\cH'_m$ for some $k$. 
 
Suppose that $\G$ fixes $\proj{\a}$. By (ii) $r(\G(F_V))=d-1$. By Corollary \ref{cor:d-1}, $\G(F_V)=F_{V'}$ for some hyperplane $V'$. Moreover, each $\ket{\a'_i}$, $i\le k$, must be real up to a phase factor. Thus, $\ket{\a^{*S}}$ is proportional to 
$\ket{\a}$ (see Definition \ref{def:S-conj}). This implies that $V'=V$ and so $\G(F_V)=F_{V'}=F_V$.

For the converse, suppose that $\G$ fixes $F_V$. Then $V$ must 
be orthogonal to $\ket{\a^{*S}}$. It follows that $\ket{\a^{*S}}$ is equal to $\ket{\a}$ up to a phase factor. Hence, 
$\G(\proj{\a})=\proj{\a^{*S}}=\proj{\a}$.
\epf

If $\ket{\a}\in\cH$ is a product vector, then obviously the 
hyperplane $\ket{\a}^\perp$ has an orthogonal basis consisting of 
product vectors. We shall prove the converse.
\bl
\label{le:prod}
Let $\ket{\a}\in\cH$ be a nonzero vector and let 
$V=\ket{\a}^\perp$. If $V$ has an orthogonal basis consisting of product vectors, then $\ket{\a}$ is a product vector.
\el
\bpf
The case $n=1$ is trivial. The bipartite case, $n=2$, is well known. Assume that $n>2$. By applying the assertion to the bipartition $\cP:=\cH_1\ox\cdots\ox\cH_k$, 
$\cQ:=\cH_{k+1}\ox\cdots\ox\cH_n$ with $1\le k<n$, we deduce that $\ket{\a}=\ket{\b}\ox\ket{\g}$ with $\ket{\b}\in\cP$ 
and $\ket{\g}\in\cQ$. Since this holds for each $k$, 
$\ket{\a}$ is a product vector.
\epf

\section{Some maximal faces in the bipartite case} \label{sec:max}

It is easy to see that there exist maximal faces of $\cS_1$ which are not induced provided that $d>6$. Indeed, there exist full states $\r\in\pars_1$, and any maximal face which contains such $\r$ is non-induced. In this section we consider only 
bipartite systems and we set $m=d_1$ and $l=d_2$. 

Given a linear map $\Phi:\End\cH_1\to\End\cH_2$, the so called 
{\em Choi matrix}, $C_\Phi\in\End\cH$, of $\Phi$ is defined by
\bea
C_\Phi:=\sum_{i,j=0}^{m-1} \ketbra{i}{j}\ox\Phi(\ketbra{i}{j}).
\eea
The map $\Phi$ is {\em positive} if $\Phi(\r)\ge0$ for all $\r\ge0$. We denote by $\cP$ the closed convex cone in 
$\Hom(\End\cH_1,\End\cH_2)$ consisting of all positive maps. We say that a point $\phi\in\cP$ is {\em exposed} if it lies on an exposed ray of the cone $\cP$.

Define the bilinear pairing 
$\langle\cdot,\cdot\rangle:\End\cH\times\Hom(\End\cH_1,\End\cH_2)\to\bC$ by 
\bea
\langle\r,\Phi\rangle:=\tr(\r^T C_\Phi),
\eea
where $T$ is the transposition map. For any subset $P\sue\cP$, 
the set 
\bea
P'=\{\r\in\cS_1:\langle\r,\phi\rangle=0,~\forall\phi\in P\}
\eea
is a face of $\cS_1$ (which may be empty). If $F$ is a face of $\cP$ then we say that $F'$ is the {\em dual face} of $F$. One defines similarly the dual of a face of $\cS_1$, which is a face of $\cP$. 

Let $\phi\in\cP$ and $\phi\ne0$. To simplify notation, we shall 
denote the face $\{\phi\}'$ also by $\phi'$. We denote by 
$P_\phi$ the set of product vectors $\ket{z}=\ket{x,y}$ such that 
$\langle\proj{z},\phi\rangle=0$ or, equivalently, 
$\phi(\proj{x})\ket{y^*}=0$. This set is important because the 
extreme points of the face $\phi'$ are exactly the pure product states $\proj{z}$ with $\ket{z}\in P_\phi$ and $||z||=1$. 
We say that $\phi$ has the {\em spanning property} if $P_\phi$ spans $\cH$. If $\phi$ is an exposed point of $\cP$ then $\phi'$ is a maximal face of $\cS_1$ 
(see \cite[Proposition 5.3]{Kye2013}).
Moreover, it follows from Corollary \ref{cor:d-1} that this maximal face is non-induced if and only if $\phi$ has the spanning property.

Let $L:\cH_2\to\cH_1$ be a nonzero linear map. Then the linear map $\phi_L:\End\cH_1\to\End\cH_2$ defined by $\phi_L(X)=L^\dag XL$ is a positive map. It has been proved recently \cite{mar2013}
that $\phi_L$ is an exposed point of $\cP$. Hence, $\phi'_L$ is a maximal face of $\cS_1$. We shall prove below that it is also induced.

Let us recall that there is a natural isomorphism, $\Psi$, of complex vector spaces $\cH=\cH_1\ox\cH_2\to\Hom(\cH_2,\cH_1)$ which sends $\ket{x}\ox\ket{y}\to\ketbra{x}{y^*}$ for each 
$\ket{x}\in\cH_1$ and $\ket{y}\in\cH_2$. While $\cH$ is already a Hilbert space, we make also $\Hom(\cH_2,\cH_1)$ into a Hilbert space by using the standard inner product 
$(X|Y):=\tr(X^\dag Y)$. Then $\Psi$ becomes an isometry.

\bpp
If $L:\cH_2\to\cH_1$ is a nonzero linear map, then the maximal face $\phi'_L$ of $\cS_1$ is induced. More precisely, if 
$\ket{\a}:=\Psi^{-1}(L)$ and $V:=\ket{\a}^\perp$ then 
$\phi'_L=F_V$.
\epp
\bpf
A product vector $\ket{z}=\ket{x,y}$ belongs to $P_{\phi_L}$ 
if and only if $\phi_L(\proj{x})\ket{y^*}=0$. This condition is equivalent to $\bra{x} L \ket{y^*}=0$, and also to 
$\tr(L^\dag \ketbra{x}{y^*})=0$, i.e., to 
$(L~|~\ketbra{x}{y^*})=0$. By applying the isometry $\Psi^{-1}$, we conclude that $\ket{x,y}\in P_{\phi_L}$ if and only if 
$\braket{\a}{x,y}=0$. Thus, $P_{\phi_L}$ is the set of all product vectors in the hyperplane $V$. Consequently, 
$\phi'_L=F_V$.
\epf

\subsection{Some $3\times3$ non-induced maximal faces}
\label{subsec:3x3max}

K.-C. Ha and S.-H, Kye have constructed recently \cite{HaKye} concrete examples of non-induced maximal faces in $3\ox3$. 
Let us mention first that they have constructed a 1-parameter family $\D_b$, $b>0$, $b\ne1$, of faces of $\cS_1$. Each $\D_b$ is a 9-dimensional simplex, the convex hull of pure product states $\proj{z_i}$, $i=1,\ldots,10$. The product vectors $\ket{z_i}$ are the normalizations of the six vectors listed in 
\cite[Eq. 10]{HaKye} and the additional four vectors listed in \cite[Eq. 12]{HaKye}. They have shown that the face $\D_b$ is the intersection of two faces $\Phi(1/b)'$ and 
$(\Phi(1/b)\circ t)'$ (see their paper for more details). Further, they have shown that 

(i) each interior point of $\D_b$ is a full state;

(ii) $(\Phi(1/b)\circ t)'=\Gamma_1( \Phi(1/b)' )$;

(iii) any product vector $\ket{z}$ for which 
$\proj{z}\in\Phi(1/b)'$ has one of the four forms listed 
in \cite[Eq. 13]{HaKye}.

To simplify the notation we set $F_b=\Phi(1/b)'$, and so by (ii) we have $(\Phi(1/b)\circ t)'=\Gamma_1(F_b)$. 
It follows from (i) that all three faces $\D_b$, $F_b$ and 
$\Gamma_1(F_b)$ are non-induced. 

It was shown in \cite{HaKye12} that the ray generated by the 
positive map $\Phi(1/b):M_3\to M_3$ (see \cite[pp. 146-7]{HaKye}) is an exposed ray of the cone of all positive maps $M_3\to M_3$. This implies (see \cite[Proposition 5.3]{Kye2013} that $F_b$ is a maximal face of $\cS_1$. Thus, the $F_b$ with $b>0$ and $b\ne1$ 
constitute a 1-parameter family of non-induced maximal faces of 
$\cS_1$.

In order to test our Conjecture \ref{cj:max}, we shall now compute $\dim F_b$, which is equal to $\dim \Gamma_1(F_b)$. 
It follows from (iii) above that the extreme points of the face 
$\Gamma_1(F_b)$ are the pure states $\proj{z}$, where $\ket{z}$ 
is the normalization of product vectors of the following four types:
\bea
\label{eq:ps1}
&&
\ket{\ps_1}=
(p,q,1) \ox (p,q,1),
\\
&&
\ket{\ps_2}=
(0,r,\sqrt{b} ) 
\ox 
(0,r\sqrt{b},1),
\\
&&
\ket{\ps_3}=
(s\sqrt{b},0,1 ) 
\ox 
(s,0,\sqrt{b} ),
\\
\label{eq:ps4}
&&
\ket{\ps_4}=
(t,\sqrt{b},0 ) 
\ox 
(t\sqrt{b},1,0 ),
\eea
where $p,q,r,s,t$ are complex numbers of modulus 1. 

Let $L_j\sue H$ denote the real span of all pure product states of type $\proj{\ps_j}$ and let $L=\sum_j L_j$. We denote by 
$L'_j$ the projection of $L_j$ on the space $T$ of upper triangular matrices with real diagonal. We define $L'$ similarly. We can write the projection of $\proj{\ps_1}$ as a linear combination of linearly independent constant matrices with 12 monomial coefficients 
$$
1,p,q,p^2,q^2,pq,p^2q^*,(pq^*)^2,pq^*,p^*q,p(q^*)^2,p^*q^2.
$$
Since this list contains only one real-valued monomial and two complex conjugate pairs, we have 
$\dim L'_0=1+2\cdot(2+7)=19$. One similarly shows that each of the spaces $L'_1,L'_2,L'_3$ has dimension 5. 
After selecting a basis for each $L'_j$, and using the fact that the union of these bases spans $L'$, a short computation (performed be Maple) shows that $\dim L'=28$. Hence, 
$\dim F_b=\dim L -1=27$. This result agrees with Conjecture \ref{cj:max}.

\subsection{Some $2\times4$ non-induced maximal faces}
\label{subsec:2x4max}

K.-C. Ha and S.-H, Kye have also constructed recently \cite{HaKye2014} a family, $\Phi=\Phi[a,b,c,d]$, of exposed positive linear maps depending on four positive real parameters $a,b,c,d$, with $ab>1$. Hence, the corresponding dual face $\Phi'$ of $\cS_1$ is maximal. Moreover, it is shown in  \cite{hk14} that each interior point of 
$\Phi'[a,b,c,d]$ is a full state and so these faces are not induced. We shall estimate their dimension in order to show 
that they do not violate Conjecture \ref{cj:max}.

The computation uses the fact \cite[Sec. 4]{hk14} that $\Phi'[a,b,c,d]$ is the convex hull of the normalizations of the pure product states $\proj{z_{\a}}$ with 
$\ket{z_\a}=\ket{x_{\a},y_{\a}}$, $\a\in\bC\cup\{\infty\}$. 
For the computation of the dimension we can ignore the case 
$\a=\infty$. The vectors $\ket{x_{\a}}$ and $\ket{y_{\a}}$ are given by
\bea
\ket{x_{\a}}=\ket{0}+\a^*\ket{1},
\quad
\ket{y_{\a}}=\sum_{j=0}^3 y_j\ket{j},
\eea
where 
\bea
&& \label{eq:prva}
y_0=g\a(1-\a),
\\
&&
y_1=\a[h-cd(\a+\a^*)+k\abs{\a}^2],
\\
&&
y_2=-e-f\abs{\a}^2,
\\
&& \label{eq:poslednja}
y_3=-\a^*(c+d\a)
\eea
The numbers $e,f,g,h,k$ are positive and defined in \cite[Eq. (1)]{hk14}, i.e., 
$$
e=\frac{ac(c+d)}{ab-1},~ f=\frac{ad(c+d)}{ab-1},~ 
g=\sqrt{acd},~ h=be-c^2,~ k=bf-d^2.
$$
Let $L\sue H$ be the real span of all pure product states $\proj{z_{\a}}$, $\a\in\bC$, and denote by $L'$ the projection of $L$ 
on the space of the upper-triangular matrices with real diagonal.  The subspace $L'$ is spanned by all matrices
\bea
\label{eq:2x4}
\sum^3_{p,q=0,p\le q}y_py_q^*\ketbra{0,p}{0,q}
+
\sum^3_{p,q=0,p\le q}y_py_q^*\abs{\a}^2\ketbra{1,p}{1,q}
+
\sum^3_{p,q=0}y_py_q^*\a\ketbra{0,p}{1,q}.
\eea

After substituting the expressions for the $y_i$ from \eqref{eq:prva}-\eqref{eq:poslednja}, we obtain a linear combination of constant matrices with monomials in $\a$ and $\a^*$ as coefficients. The monomials that occur are 
$(\a^*)^q$ with $q=0,1,2,3$, 
$\a(\a^*)^q$ with $q=1,2,3,4$, 
$\a^p(\a^*)^q$ with  $p=2,3,4$ and $q=0,1,2,3,4$, 
and $\a^5(\a^*)^3$. 
Among these 24 monomials there are five real valued, $|\a|^{2p}$ with $p=0,\ldots,4$, eight complex conjugate pairs, and three 
complex-valued which are not paired with their conjugates. 
Hence, $\dim L'\le 5+2(8+3)=27$ and $\dim \Phi'\le 26$.

\section{Conclusion and discussion} \label{sec:conclusion}

We have computed the dimension of induced maximal faces $F_V$, 
$V$ a hyperplane, of the compact convex set of separable states in any finite-dimensionsl multipartite quantum system $\cH$. As a consequence, we obtain that the rhs of Eq. \eqref{eq:Max} in Conjecture \ref{cj:max} is equal to $d(d-2)$, where $d$ is the dimension of $\cH$. However the lhs of this equation remains unknown. Since every non-induced face is proper, we can reformulate Conjecture \ref{cj:max} as follows: 
For any non-induced face $F$ of $\cS_1$ we have 
$\dim F\le d(d-2)$.

We have characterized the genuine entanglement contained in a pure state $\ket{\a}$ in terms of the induced maximal face $F_V$, where $V=\ket{\a}^\perp$. Indeed, we have shown that $\ket{\a}$ is genuinely entangled if and only if 
$\dim F_V=d(d-2)$. It would be of interest to generalize our results to the genuine entanglement of mixed states. 

In the bipartite case, $n=2$, there is a duality between the cone $\cS$ of non-normalized separable states and the cone $\cP$ of positive linear maps $\End\cH_1\to\End\cH_2$. If $\phi\in\cP$ 
generates 
an exposed ray of $\cP$ (in that case we say that $\phi$ is an exposed point of $\cP$) then the dual face of $\cS$ is maximal, 
and its intersection with $\cS_1$ is a maximal face which we denote by $\phi'$. If $L:\cH_2\to\cH_1$ is a nonzero linear map, 
it has been shown \cite{mar2013} that the positive map 
$\phi_L:\End\cH_1\to\End\cH_2$ defined by $\phi_L(X)=L^\dag XL$ is an exposed point of $\cP$. We show that the corresponding maximal faces $\phi'_L$ are exactly the maximal faces $F_V$ with 
$V\subseteq\cH$ a hyperplane.

The non-induced maximal faces of $\cS_1$ are much harder to
construct. We have examined two such families, one in $3\ox3$ and one in $2\ox4$. In both cases we have verified that their dimensions are smaller than $d(d-2)$ which agrees with 
Conjecture \ref{cj:max}.
 
Recently multipartite genuine entanglement has been related to the quantum marginal problem via local and global properties of quantum states \cite{cgm14}. As pointed out in \eqref{eq:ge-decomp}, any pure quantum state can be expressed as the tensor product of genuinely entangled states. In entanglement theory, for any entanglement measure $E$ we have $E(\ox_i \ket{\ps_i})=\sum_i E(\ket{\ps_i})$. All these facts imply that characterizing genuine entanglement is a fundamental problem in quantum information.

\section*{Acknowledgments}

We thank an anonymous referee for pointing out that $d(d-2)$ is 
the expected dimension for maximal faces of $\cS_1$.
LC was partially supported by the Fundamental Research Funds for the Central Universities (Grant No. 30426401 and Grant No. 30458601). The work started when LC was also supported in part by the Singapore National Research Foundation under NRF Grant No. NRF-NRFF2013-01. DD was supported in part by an NSERC Discovery Grant.

\section*{Conflict of Interest Statement}

Conflict of Interest: The authors declare that they have no conflict of interest.

\end{document}